\documentclass[onecolumn, floatfix, preprintnumbers, eqsecnum, letterpaper, superscriptaddress, nofootinbib]{revtex4}
\usepackage{graphicx}
\usepackage{microtype}
\usepackage{amsmath}
\usepackage{amssymb}
\usepackage{subfigure}
\usepackage{hyperref}
\usepackage{url}
\usepackage{xcolor}
\usepackage{color}
\usepackage{mathrsfs}
\usepackage{calrsfs}
\usepackage{amsfonts}
\usepackage{eufrak}
\usepackage{tabularx}
\usepackage{eucal}
\usepackage{latexsym}
\usepackage{ragged2e}
\usepackage{epsfig}
\usepackage{textcomp}

\usepackage{caption}
\usepackage{subfigure}

\DeclareCaptionJustification{justified}{\leftskip=0pt \rightskip=0pt \parfillskip=0pt plus 1fil}
\definecolor{vividviolet}{rgb}{0.62, 0.0, 1.0}
\definecolor{amaranth}{rgb}{0.9, 0.17, 0.31}
\definecolor{palatinateblue}{rgb}{0.15, 0.23, 0.89}
\definecolor{brightpink}{rgb}{1.0, 0.0, 0.5}
\definecolor{cornflowerblue}{rgb}{0.39, 0.58, 0.93}
\definecolor{deepcarminepink}{rgb}{0.94, 0.19, 0.22}
\definecolor{radicalred}{rgb}{1.0, 0.21, 0.37}

\hypersetup{ linktoc=all,
    colorlinks, linkcolor={palatinateblue},
    citecolor={brightpink}, urlcolor={amaranth}
}

\graphicspath{{Images/}}

\graphicspath{{Images/}}

\def\@fnsymbol#1{\ensuremath{\ifcase#1\or \ddagger \or  $\textleaf$  \or \dagger
\else\@ctrerr\fi}}%

\makeatother

\def\sideremark#1{\ifvmode\leavevmode\fi\vadjust{\vbox to0pt{\vss
 \hbox to 0pt{\hskip\hsize\hskip1em
 \vbox{\hsize1.3cm\tiny\raggedright\pretolerance10000
 \noindent #1\hfill}\hss}\vbox to8pt{\vfil}\vss}}}%

\def\beq{\begin{equation}}
\def\eeq{\end{equation}}

\begin{document}

\title{Heat Capacity and the Violation of Scaling Laws in Gravitational System}

\author{Shi-Bei Kong}
\email{shibeikong@ecut.edu.cn}
\affiliation{School of Science, East China University of Technology, Nanchang 330013, Jiangxi, China}

\begin{abstract}

In this paper, we examine the scaling laws in gravitational system from the perspective of free energy landscape
and the scaling hypothesis. It has been found that for some special black holes, their critical exponents $(0,1,2,3)$ are beyond the mean field theory, 
and more surprisingly violate the scaling laws. We find that the main reason for the violation of the scaling laws 
is that the heat capacity at constant volume $C_V$ is 0, so the critical exponent $\alpha$ is often treated as 0,
which can not be derived from the scaling hypothesis.
We also find that there is a symmetry violation for the two coexistence states $\omega_l$ and $\omega_s$.

\end{abstract}

\maketitle

\section{Introduction}

Phase transitions and critical phenomena of the van der Waals-type in gravitational systems have been widely and deeply investigated 
in recent years \cite{Kubiznak:2012wp,Wei:2012ui,Cai:2013qga,Xu:2015rfa}.
In the extended phase space of black hole thermodynamics \cite{Gunasekaran:2012dq,Karch:2015rpa,Kubiznak:2016qmn,Mancilla:2024spp}, 
the cosmological constant is treated as the thermodynamic pressure, and in the thermodynamics of the FRW(Friedmann-Robertson-Walker) 
universe \cite{Cai:2005ra,Cai:2006rs,Abdusattar:2021wfv,Kong:2021dqd,Kong:2022xny,Abdusattar:2023hlj,Chu:2025zuz,Feng:2024zor},
the work density plays the role of the thermodynamic pressure, which satisfies a van der Waals-like equation of state $P=P(V,T)$. 
Various equations of state for black holes and the FRW universe have critical points where second 
order phase transitions take place. Therefore, near the critical point, critical exponents can be defined and calculated \cite{Pelissetto:2000ek}.
The most common critical exponents are ($\alpha, \beta, \gamma, \delta$), which are related with the heat capacity at constant volume $C_V$,
order parameter $\eta$, isothermal compressibility $\kappa_T$, and the shape of isotherm along the critical temperature respectively.
For nearly all the cases, the exponents are the same with those in the van der Waals-system or mean field theory, i.e. $(0,1/2,1,3)$, 
so they satisfy the scaling laws \cite{Pelissetto:2000ek}. However, non-standard exponents $(0,1,2,3)$ are also found for some special black holes,
e.g. Lovelock black holes \cite{Dolan:2014vba,Hennigar:2016ekz}, hairy black holes in cubic quasi-topological gravity \cite{Dykaar:2017mba},
and black holes with quantum anomaly \cite{Hu:2024ldp}, etc. 
The special critical points with non-standard exponents $(0,1,2,3)$ are called isolated critical points \cite{Dolan:2014vba} and their thermodynamic topology 
has also been studied \cite{Wu:2025xxo}. They are beyond the mean field theory and even seem to violate the scaling laws. 
This poses a serious question on the universality of the scaling laws near the critical point, which needs to be verified and explained correctly. 

In this work, we start from the scaling hypothesis and show that the critical exponents $(0,1,2,3)$ indeed violate the scaling laws, and what is more,
this violation is mainly caused by the first critical exponent $\alpha=0$ associated with $C_V$. To be specific, 
we first assume that the scaling hypothesis is satisfied, and find that the critical exponents
should be $(-2,1,2,3)$ rather than $(0,1,2,3)$, so the latter combination indeed violates the scaling hypothesis and the scaling laws.
It is very obvious that this violation is mainly caused by the first critical exponent $\alpha$ 
rather than the other three exponents ($\beta, \gamma, \delta$). For black holes including these special black holes \cite{Dolan:2014vba,Hennigar:2016ekz,Dykaar:2017mba,Hu:2024ldp}, 
$C_V$ is zero, and the critical exponent $\alpha$ is typically regarded as 0 instead of -2. 
Therefore, the violation of the scaling law stems from the special behavior of the black hole heat capacity. 
  
This paper is organized in the following way. In Sec.II, we show how to get the critical exponents for the usual van der Waals system
from the scaling hypothesis. In Sec.III, we show how to get the critical exponents for these special black holes from the scaling hypothesis
and why the scaling laws are violated. In Sec.IV, we make the conclusions and some discussions. We use natural units with $c=\hbar=G=1$.

\section{Critical Exponents of van der Waals System}

Near the critical point, the four critical exponents can be defined as \cite{Pelissetto:2000ek}:
\begin{eqnarray}
C_{V}&=&T{\left(\frac{\partial S}{\partial T}\right)}_{V}\sim \left| t \right|^{-\alpha} , \label{alpha}
\\
\eta&=&\rho_{l}-\rho_{g}\sim v_{g}-v_{l}\sim|t|^{\beta}  , \label{beta}
\\
\kappa_{T}&=&-\frac{1}{V}{\left(\frac{\partial V}{\partial P}\right)}_{T}\sim\left| t
\right|^{-\gamma},
\\
\left| P-P_{c} \right|&\sim&\left| \rho-\rho_{c} \right|^{\delta},
\end{eqnarray}
where `$g$' and `$l$' stand for the `gas' and `liquid' states respectively.

The van der Waals equation is

\begin{alignat}{1}
(P+\frac{a}{v^2})(v-b)=k T,  \label{vdW}
\end{alignat}
where $v:=V/N$, and it has a critical point at
\begin{alignat}{1}
P_c=\frac{a}{27b^2}, \quad v_c=3b, \quad T_c=\frac{8a}{27kb}. 
\end{alignat}

The Gibbs free energy of the vdW system is

\begin{alignat}{1}
G(P,T,v)=-k T\left(1+\ln\left[\frac{(v-b)T^{\frac{3}{2}}}{\Phi}\right]\right)-\frac{a}{v}+P v,
\end{alignat}
which can be expanded around the critical point with $t\rightarrow 0$ and $\omega\rightarrow 0$ as
\begin{alignat}{1}
G(P,T,v)=G_0(P,T)+\frac{a}{9b}(p-4t)\omega+\frac{a}{3b}t\omega^2-\frac{a}{3b}t\omega^3+\frac{3a}{8b}(1+9t)\omega^4
+\mathcal{O}(\omega^5,t\omega^5),
\end{alignat}
where $p:=P/P_c-1, t:=T/T_c-1, \omega:=v/v_c-1$. For the van der Waals system, $a$ and $b$ are not zero, so the $t\omega^3$ and $t\omega^4$
terms are less dominant than $t\omega$ and $t\omega^2$ near the critical point, thus they can also be dropped as well as high order terms. 
Therefore, we get the free energy in the Landau form
\begin{alignat}{1}
G(P,T,v)=G_0(P,T)+\frac{a}{9b}h\omega+\frac{a}{3b}t\omega^2+\frac{3a}{8b}\omega^4+\mathcal{O}(\omega^5,t\omega^3), \label{LFE}
\end{alignat}
where $h:=p-4 t$ is the `external' field, which is consistent with (153.1) in the classic book \cite{Landau:1980mil}.

Varying the Landau free energy (\ref{LFE}) with the parameter $\omega$, one can get the equation of state around the critical point
\begin{alignat}{1}
h=-6t\omega-\frac{3}{2}\omega^3+\mathcal{O}(\omega^4,t\omega^2), \label{eosc}
\end{alignat}
which is also consistent with the direct expansion of the vdW equation (\ref{vdW}) around the critical point.\footnote{One can also start from
(\ref{eosc}) and recover the corresponding Landau free energy (\ref{LFE}) as did in \cite{Landau:1980mil}.}

The gas and liquid states have the same free energy and control parameters, i.e. $G(T,P,v_l)=G(T,P,v_g),h_l=h_g=h^*$, so
\begin{alignat}{1}
\frac{1}{9}h^*\omega_l+\frac{1}{3}t\omega_l^2+\frac{3}{8}\omega_l^4
=&\frac{1}{9}h^*\omega_g+\frac{1}{3}t\omega_g^2+\frac{3}{8}\omega_g^4,
\\
-6t\omega_l-\frac{3}{2}\omega_l^3=&-6t\omega_g-\frac{3}{2}\omega_g^3=h^*,
\end{alignat}
which are equivalent to the Maxwell's equal area law. From the above two equations, one can get
\begin{alignat}{1}
\omega_g=2\sqrt{-t}, \quad \omega_l=-2\sqrt{-t}, \label{wt}
\end{alignat}
which means that the second exponent $\beta$ is $1/2$. The results show that the absolute values of the two states $\omega_g$ and $\omega_l$ are the same,
so they are symmetric about the critical point with $\omega_c=v_c/v_c-1=0.$
From the results and (\ref{eosc}), we can also see that the external field $h$ is zero for the two coexistence states, which means that the $h=0$ line is the equal-area line, and the phase transition can be regarded as a spontaneous one.

In the following, we discuss the scaling hypothesis. At first, we should write the free energy as a function of $p$ and $t$ only.
From (\ref{wt}), one can see that $\omega$ scales as $(-t)^{1/2}$, so it can be written as $\omega=c(-t)^{1/2}$, 
and the free energy can be written as
\begin{alignat}{1}
G(p,t)=G_0(p,t)+Ah(-t)^{1/2}+Bt^2, \label{Gpt}
\end{alignat}
where $A\equiv ac/(9b),B\equiv3a c^4/(8b)-ac^2/(3b)\neq 0$.
The scaling hypothesis means that the exotic part of the free energy, i.e. $G(p,t)-G_0(p,t)=f(t,h)$, satisfies
\begin{alignat}{1}
\lambda f(t,h)=f(\lambda^pt,\lambda^qh), \label{sh}
\end{alignat}
and the four critical exponents can be written by $p$ and $q$:
\begin{alignat}{1}
\alpha=2-\frac{1}{p}, \quad \beta=\frac{1-q}{p}, \quad \gamma=\frac{2q-1}{p}, \quad \delta=\frac{q}{1-q}.
\end{alignat}
They satisfy the four scaling laws:
\begin{alignat}{1}
\alpha+2\beta+\gamma=&2,\quad \alpha+\beta(1+\delta)=2, \quad \gamma(1+\delta)=(2-\alpha)(\delta-1),\quad \gamma=\beta(\delta-1). \label{sl}
\end{alignat}
Therefore, apply the scaling hypothesis to (\ref{Gpt}), one can get
\begin{alignat}{1}
\lambda[Ah(-t)^{1/2}+Bt^2]=A(\lambda^qh)(-\lambda^pt)^{1/2}+B(\lambda^pt)^2,
\end{alignat}
which results to
\begin{alignat}{1}
1=q+\frac{p}{2}, \quad 1=2p,
\end{alignat}
so
\begin{alignat}{1}
p=\frac{1}{2}, \quad  q=\frac{3}{4},
\end{alignat}
and the four critical exponents are
\begin{alignat}{1}
\alpha=0, \quad \beta=\frac{1}{2}, \quad \gamma=1, \quad \delta=3.
\end{alignat}
This is exactly the results from the mean field theory and of course satisfy the four scaling laws (\ref{sl}).
It should be noted that a more common method is to replace $\omega$ with $c h^{1/3}$, but the results are the same.

\section{Critical Exponents of Black Holes}

For most black holes with a vdW-like phase transition, one can also get their Landau free energy similar to (\ref{LFE}), which can be generally
written as
\begin{alignat}{1}
G(P,T,V)=G_0(P,T)+(p-At)\omega-\frac{B}{2} t\omega^2-\frac{C}{4}\omega^4+\mathcal{O}(\omega^5,t\omega^3), \label{expansion}
\end{alignat}
and the corresponding equation of state is
\begin{alignat}{1}
p=At+B t\omega+C\omega^3+\mathcal{O}(\omega^4,t\omega^2),
\end{alignat}
where $p:=P/P_c-1,t:=T/T_c-1,\omega:=V/V_c-1$, and $A,B,C$ are non-zero coefficients.

However, for some special black holes, such as the black hole with quantum anomaly \cite{Cai:2009ua,Cai:2014jea,Hu:2024ldp} and the 4d charged Gauss-Bonnet black hole in AdS space \cite{Wei:2020poh}, this form (\ref{expansion}) is not suitable under certain conditions. Take the black hole with quantum anomaly as an example,
its equation of state is
\begin{eqnarray}
P=\frac{T}{2 r_h}-\frac{1}{8 \pi  r_h^2}-\frac{2\alpha_c  T}{r_h^3}+\frac{Q^2-2\alpha_c}{8 \pi  r_h^4},
\end{eqnarray}
and its Gibbs free energy is
\begin{alignat}{1}
G=\frac{1}{6r_h}\left[-6\alpha_c+r_h^2(8\pi Pr_h^2-6\pi T r_h+3)+48\pi\alpha_c T r_h\ln\left(\frac{r_h}{\sqrt{2|\alpha_c|}}\right)+3Q^2 \right].
\end{alignat}
At first we expand this free energy as (\ref{expansion}) and the coefficients are
\begin{alignat}{1}
A=\frac{4(3Q^2-4\alpha_c+K)}{9Q^2-22\alpha_c}, \quad B=\frac{6Q^2-48\alpha_c-10K}{27Q^2-66\alpha_c},
\quad C=\frac{40\alpha_c-15Q^2+K}{27(9Q^2-22\alpha_c)},
\end{alignat}
where $K=\sqrt{192\alpha_c^2+9Q^4-96\alpha_cQ^2}$. It can be seen that if the condition $Q^2-8\alpha_c=0$ is satisfied, one has $B=0$
and the $t\omega^2$ term vanishes, so this form (\ref{expansion}) is not appropriate and 
one should add higher order term $t\omega^3$ in the free energy. Therefore,
the Landau free energy should be written as
\begin{alignat}{1}
G(P,T,V)=G_0(P,T)+(p-At)\omega-\frac{C}{4}\omega^4-\frac{D}{3}t\omega^3+\mathcal{O}(\omega^5,t\omega^4), \label{fe}
\end{alignat}
and the equation of state is
\begin{alignat}{1}
p=At+C\omega^3+Dt\omega^2+\mathcal{O}(\omega^4,t\omega^3).
\end{alignat}
Similar to the previous section, i.e. $G(T,P,v_l)=G(T,P,v_s),h_l=h_s=h^*$, one can get
\begin{alignat}{1}
\omega_l=\frac{\sqrt{3}-1}{3}\left|\frac{D}{C}t\right|, \quad \omega_s=-\frac{\sqrt{3}+1}{3}\left|\frac{D}{C}t\right|.
\end{alignat}
In this case, the above results or equal area law do not result to a zero external field, i.e. $h=p-At\neq0$. 
What's more, the absolute values of $\omega_l$ and $\omega_s$ are not the same, which is different with the van der Waals case
(\ref{wt}). Therefore, if $B=0$, the two states $\omega_l$ and $\omega_s$ are not symmetric about the critical point with $\omega_c=V_c/V_c-1=0$.
This could be regarded as a symmetry violation. The results are closely related with the order parameter as $\eta\propto v_l-v_s\propto\omega_l-\omega_s$ (\ref{beta}), so one can read out $\beta=1$ instead of $\beta=1/2$ in the van der Waals system. 
From the results, one can also see that $\omega$ scales as $t$, so it can be written as $\omega=\tilde{D}t$ in the following calculations. 

Now we discuss the scaling hypothesis. The exotic part \cite{Landau:1980mil} of the free energy (\ref{fe}) can be written as
\begin{alignat}{1}
f(t,h)\equiv G(P,T,V)-G_0(P,T)=&(p-At)\omega-\frac{C}{4}\omega^4-\frac{D}{3}t\omega^3+\mathcal{O}(\omega^5,t\omega^4)
\nonumber \\
=&\tilde{D}h t-\left(\frac{C\tilde{D}}{4}+\frac{D}{3}\right)\tilde{D}^3t^4+\mathcal{O}(\omega^5,t\omega^4),
\end{alignat}
where $h=p-At, \omega=\tilde{D}t$ are used. Therefore, if it satisfies the scaling hypothesis (\ref{sh}), one can get
\begin{alignat}{1}
p=\frac{1}{4}, \quad  q=\frac{3}{4},
\end{alignat}
which means
\begin{alignat}{1}
\alpha=-2, \quad \beta=1, \quad \gamma=2, \quad \delta=3.
\end{alignat}
This is beyond the mean field theory but satisfies the scaling laws.
Compared with the four non-stardard exponents ($\alpha=0, \beta=1, \gamma=2, \delta=3$), one finds that only
the first one is different, so the violation of the scaling laws is mainly caused by the critical exponent $\alpha$. This exponent is associated with the heat capacity at constant (thermodynamic) volume $C_V$, which is calculated as
\begin{alignat}{1}
C_V=\left(\frac{dQ}{dT}\right)_V=\left(\frac{TdS}{dT}\right)_V=T\left(\frac{\partial S}{\partial T}\right)_V,
\end{alignat}
where $S=A/4-4\pi\alpha_c\ln(A/A_0)$ \cite{Hu:2024ldp} is the black hole entropy\footnote{Other special black holes may have different forms of entropy,
but are usually functions of $A$ or $r_h$, so the conclusions are the same.} with $A=4\pi r_h^2$ and $V=4\pi r_h^3/3$. 
Obviously, if the (thermodynamic) volume $V$ is fixed, $r_h$ does not change, so consequently the entropy $S$ does not change thus $C_V=0$. 
According to the definition (\ref{alpha}) of the critical exponent $\alpha$, one can see that $C_V=0$ is independent of $t$, so $\alpha$ should be 0 \cite{Kubiznak:2012wp}. Therefore, a zero heat capacity at constant volume is the key reason that scaling laws are not satisfied for these black holes. 
$C_V=0$ is a special thermodynamic feature of almost all the black holes, which is rarely seen in other thermodynamic systems.
However, it should be noted that $C_V=0$ does not always result to the violation of scaling laws because most black holes with $C_V=0$
also satisfy the scaling laws.

\section{Conclusions and Discussions}

In this paper, we show that the strange new critical exponents ($\alpha=0, \beta=1, \gamma=2, \delta=3$) of some special black holes violate the scaling laws in a more apparent and direct way, i.e. the scaling hypothesis. To be specific, if the scaling laws or scaling hypothesis are satisfied, 
the critical exponents should be ($\alpha=-2, \beta=1, \gamma=2, \delta=3$) for these black holes,
so ($\alpha=0, \beta=1, \gamma=2, \delta=3$) does not satisfy the scaling laws.
Apparently, only the first exponent $\alpha$ is different, so the violation 
of the scaling laws is mainly caused by $\alpha=0$, which is resulted from $C_V=0$ being independent of $t$. 

However, it should be noted that most black holes with $C_V=0$ do not violate the scaling laws,
and only those black holes that satisfy certain conditions can violate the scaling laws. 
These black holes are usually derived from modified gravity or quantum effect, so modified gravity or quantum effect is also an important reason.
The violation of scaling laws may also appear in the phase transitions of the FRW universe and especially under modified gravity.

\section*{Acknowledgment}

I would like to thank Haximjan Abdusattar, Yu-Sen An, Liang Cai, Yue Chu, Ya-Peng Hu, Xiao Liang, Shao-Wen Wei, Yi-Hao Yin,  Wen-Long You, Hong-Sheng Zhang et al. for helpful discussions. I also thank the reviewer for his/her questions and comments. This work is supported by National Natural Science Foundation of China (NSFC) under No.12465011
and East China University of Technology (ECUT) under DHBK2023002.

\end{document}